\documentclass[aps,nofootinbib,longbibliography,notitlepage,superscriptaddress,twocolumn,10pt,prd]{revtex4-2}

\usepackage[utf8]{inputenc}
\usepackage[dvips]{graphicx} %
\usepackage{graphicx,amsmath,amsfonts,amssymb,slashed,float,hyperref}
\usepackage{bbold,wasysym}
\usepackage{array,multirow}
\usepackage[usenames,dvipsnames]{xcolor} 
\usepackage{soul}

\definecolor{blop}{RGB}{184, 24, 199}

\definecolor{ForestGreen}{RGB}{36,179,0}

\definecolor{KindaGreen}{RGB}{57, 192, 237}

\newcommand{\LCDM}{$\Lambda$CDM}
\newcommand{\Mpl}{{M_{\rm Pl}}}

\newcommand{\dm}{{\rm dm}}
\newcommand{\ede}{{\rm ede}}
\hypersetup{
	colorlinks=true,
	linkcolor=red,
	citecolor=blue,
}

\begin{document}
\title{Chameleon Early Dark Energy and the Hubble Tension}

\author{Tanvi Karwal}
\author{Marco Raveri}
\author{Bhuvnesh Jain}
\author{Justin Khoury}
\author{Mark Trodden}

\affiliation{Center for Particle Cosmology, Department of Physics and Astronomy, University of Pennsylvania, Philadelphia, PA 19104, USA}

\begin{abstract}
    Early dark energy (EDE) offers a particularly interesting theoretical approach to the Hubble tension, albeit one that introduces its own set of challenges, including a new `why then' problem related to the EDE injection time at matter-radiation equality, and a mild worsening of the large-scale structure (LSS) tension. 
    Both these challenges center on the properties of dark matter, which becomes the dominant component of the Universe at EDE injection and is also responsible for seeding LSS.
    Motivated by this, we present the potential of couplings between EDE and dark matter to address these challenges, focusing on a mechanism similar to chameleon dark energy theories, deeming this chameleon early dark energy (CEDE). 
    We present relevant background and perturbation equations and study the dynamics of the case of a quartic scalar potential and an exponential coupling. 
\end{abstract}

\maketitle

\section{Introduction} \label{Sec:Intro}

The $\Lambda$CDM model has accumulated an impressive amount of support from cosmological data. 
Yet, numerous persistent and fascinating anomalies remain, presenting the tantalizing possibility of new fundamental physics. 
Perhaps the most pressing of these is the discrepancy between measurements of the Hubble constant $H_0$ made using different observables at vastly different redshifts. 
Specifically, estimations of $H_0$ based on early-universe data \cite{Addison:2017fdm,Aghanim:2018eyx,Addison:2021amj} are consistently and significantly lower than measurements using late-universe data \cite{Riess:2019cxk,Pesce:2020xfe,Huang:2019yhh,Freedman:2019jwv,Yuan:2019npk,Freedman:2020dne}, with the resulting conundrum known as the {\it Hubble tension} \cite{Verde:2019ivm,Freedman:2017yms}.
Although the possibility remains that the source of this discrepancy lies in unresolved systematics \cite{Rigault:2018qlp,NearbySupernovaFactory:2018jea,Birrer:2020tax,1981089,Dainotti:2021pqg}, the alternative is far more exciting - that the Hubble tension indicates new physics which will have other quantifiable, observable signatures. 

Theorists have mounted a huge collective effort to understand this anomaly, with approaches naturally falling into two categories: those in which the physics of the late universe is altered [see \cite{DiValentino:2020zio} for a review], and others in which new physics is introduced at early epochs \cite{Karwal:2016vyq,Poulin:2018dzj,Poulin:2018cxd,Lin:2019qug,Agrawal:2019lmo,Berghaus:2019cls,Sakstein:2019fmf,Kreisch:2019yzn,Ghosh:2019tab,Smith:2019ihp,Pandey:2019plg,Niedermann:2019olb,Alexander:2019rsc,Ballardini:2020iws,Braglia:2020auw,Allali:2021azp}. 
Late-universe modifications have found severe challenges due to competing constraints from numerous independent datasets at low redshifts \cite{Raveri:2019mxg,Benevento:2020fev,Alestas:2021xes,Umilta:2015cta} (but see \cite{Desmond2019,Sakstein2019} for a partially successful late-universe modification).
Early-universe modifications, while also facing significant constraints, have been somewhat more successful \cite{Bernal:2016gxb,Aylor:2018drw,Knox:2019rjx,Evslin:2017qdn,Vagnozzi:2021gjh,Kreisch:2019yzn}, with early dark energy (EDE) holding the most promise \cite{Karwal:2016vyq,Poulin:2018cxd,Lin:2019qug,Agrawal:2019lmo,Smith:2020rxx}. 

EDE scenarios, first introduced in Ref.~\cite{Kamionkowski:2014zda}, add a new scalar-field component to the Universe which is insignificantly subdominant at all times except for a redshift-localized contribution close to matter-radiation equality. 
From a particle-physics perspective, a serious drawback of such a field is the excessive fine-tuning of its parameters, including exceptionally small masses, required to achieve kination at the right moment in cosmic history. 
A natural next step is to consider whether EDE might be coupled to other components, such that its dynamics are triggered by some other, pre-existing physics concurrent with the epoch of equality. 
This can be achieved, for example, by coupling EDE to neutrinos which coincidentally decouple around $z_{\rm eq}$ ~\cite{Sakstein:2019fmf,CarrilloGonzalez:2020oac}.

Couplings between fields within the dark sector of cosmology have been invoked in many other contexts \cite{Benisty:2021cmq}, most notably to explain late-time dark energy \cite{Brax:2004qh,Boriero:2015loa,Cai:2021wgv}. 
Here, we explore an EDE field that conformally couples to dark matter, inspired by chameleon models of dark energy, 
designating this cosmology `chameleon EDE' or CEDE, 
such that its dynamics are triggered by dark matter becoming the dominant component of the Universe close to matter-radiation equality at $z_{\rm eq}$. 
Such a coupling essentially translates to a modification of the effective potential felt by the scalar, and a modulation of the dark matter mass. 

The introduction of this coupling address not only the fine-tuning of the EDE injection time, but also various other criticisms of EDE models. 
A scalar-field model of EDE that is especially successful at improving the Hubble tension and the fit to cosmological data has $V(\phi) \sim (1 - \cos \phi)^n$ for integer $n$ \cite{Smith:2019ihp}, which is difficult to theoretically motivate. 
Moreover, the solution to the Hubble tension with the above flavor of EDE lies at the edges or beyond the parameter space allowed by physical priors on the theoretical parameters \cite{Hill:2020osr}. 
Here, we aim to address both these criticisms by constructing a model with a well-motivated potential, and scanning the parameter space on both theoretical and phenomenological parameters to verify consistency across parameter bases. 

Lastly, we are motivated to explore this EDE-dark matter coupling to search for combined solutions to the Hubble and large-scale structure (LSS) tensions. 
The Hubble tension is not alone in puzzling cosmologists.
A different tension, though so far not highly statistically significant, exists between the CMB-$\Lambda$CDM expectation versus local-universe constraints on the amplitude of matter density fluctuations in the late universe, parameterised by $S_8$ \cite{DES:2017lqs,Hildebrandt:2018yau,HSC:2018mja,KiDS:2020vjq,Hildebrandt:2017qln,Abbott:2021bzy}.
Although this LSS tension has a lower significance than the Hubble tension, it has persisted across different measures of LSS (redshift space distortions, weak lensing, galaxy clusters) as measurement precision has increased. 
Unfortunately, EDE scenarios tend to worsen this tension while relieving the Hubble tension \cite{Hill:2020osr,Ivanov:2020ril,Murgia:2020ryi,DAmico:2020ods}. 
In coupling EDE to dark matter, we modify the evolution of dark matter, with the possibility of  simultaneously easing the Hubble and LSS tensions. 

In this paper, we present the background and perturbative evolution of the components of such a cosmology. 
We will pursue data constraints on chameleon EDE (CEDE) models in future work. 
We present the background evolution of a universe with a scalar field conformally coupled to dark matter in Section~\ref{Sec:Background} and its perturbations dynamics in Section~\ref{sec:Perturbations}. 
In Section~\ref{Sec:InitialConditions}, we briefly discuss initial conditions for the background and perturbations. 
We explore a specific realization of CEDE in Section~\ref{Sec:CEDEmodels}, making specific choices for the scalar field's native potential and the form of its coupling to DM and illustrate its impact on CMB data in Section~\ref{Sec:BetaResults}.
Finally, we conclude in Section~\ref{Sec:Conclusions}, looking to the future of the Hubble tension and EDE. \footnote{A previous version of this article contained an error in Eq.~\eqref{Eq:dm_continuity_pertb} (missing factor of $A$ in the denominator of the $A^2_{,\phi}$ term). The code the analysis was based on contained another error in Eq.~\eqref{Eq:KG_pertb} (flipped signs on the right-hand side). These have since been corrected and the analysis updated to reflect these changes, removing some of the results based on erroneous code. }

\section{Background evolution} \label{Sec:Background}
We begin with the chameleon action $S$:
\begin{align} \label{Eq:ChameleonAction}
    S = & \int {\rm d}^4 x\sqrt{-g} \left[ \frac{\Mpl}{2}\mathcal{R} -\frac{1}{2}(\nabla \phi)^2 -V(\phi) \right] \nonumber \\ 
    & +S_{\dm}[\phi_\dm, \tilde{g}_{\mu\nu} ] +S_{\rm m}[\phi_{\rm m}, g_{\mu\nu} ] \,,
\end{align}
where $m$ denotes all standard matter species, baryons, photons and neutrinos, $\Mpl$ is the reduced Planck mass, 
$g$ is the determinant of the metric $g_{\mu \nu}$ 
and the two metrics, $g_{\mu \nu}$ and $\tilde{g}_{\mu \nu}$, are related through some arbitrary function $A(\phi)$ of the scalar $\phi$ by 
\begin{align} \label{Eq:ConformalMetric}
    \tilde{g}_{\mu\nu} = A^2(\phi) g_{\mu\nu} \,.
\end{align}
The two metrics define two reference frames, the Einstein frame with the metric $g_{\mu \nu}$ in which photons and baryons move on geodesics but DM motion is influenced by acceleration due to the scalar field,
and the DM or Jordan frame with metric $\tilde{g}_{\mu \nu}$ in which DM moves on geodesics but photons and baryons are accelerated.
We indicate quantities defined in the Jordan frame with a tilde and quantities without a tilde are assumed to be defined in the Einstein frame.
As the choice of reference frame where we perform calculations does not impact physical results, we generally work and solve equations in the Einstein frame for convenience.
The uncoupled scalar field model is recovered in the limit $A(\phi) \rightarrow 1$.

The two stress-energy tensors for regular matter and DM are defined in the Einstein frame as:
\begin{align}
    T_{\mu\nu}^{\rm m} = -\frac{2}{\sqrt{-g}}\frac{\delta \mathcal{L}_{\rm m}}{\delta g^{\mu\nu}} \,,
\end{align}
and
\begin{align}
    T_{\mu\nu}^\dm = - A^6(\phi) \frac{2}{\sqrt{-\tilde{g}}}\frac{\delta \mathcal{L}_\dm}{\delta \tilde{g}^{\mu\nu}} 
    = A^6(\phi) \tilde{T}_{\mu\nu}^\dm \,,
\end{align}
clarifying the relation between the DM stress-energy tensor in the two conformal frames.

We assume that in their respective geodesic frames, different matter species have a stress-energy tensor of a perfect fluid.
In particular, for the DM component we have 
\begin{align}
    \tilde{T}_{\mu\nu} = (\tilde{\rho} +\tilde{P})\tilde{u}_\mu\tilde{u}_\nu +\tilde{P}\tilde{g}_{\mu\nu} \,.
\end{align}
Note that, since $\tilde{u}_\mu$ is a time-like geodesic of $\tilde{g}$, then $\tilde{u}_\mu\tilde{u}_\nu \tilde{g}^{\mu\nu} = -1$.
The physical density and pressure of DM are defined in the rest frame of the DM fluid, which is accelerated in the baryon frame.
Both stress-energy tensors are covariantly conserved with respect to their geodesic frame metric 
in the absence of non-gravitational interactions:
\begin{align}
    \nabla_\mu T^{\mu\nu}_{\rm m} = 0 \hspace{0.5cm}\mbox{and}\hspace{0.5cm} \tilde{\nabla}_\mu \tilde{T}^{\mu\nu}_\dm = 0 \,.
\end{align}
From this, we see that it is convenient to work in the Einstein frame where baryons move on geodesics, since we can more easily quantify their Thompson scattering with photons.

At the background level in the Einstein frame, the Friedmann equation reads
\begin{align}
    3 \mathcal{H}^2 \Mpl^2
    = &\frac{1}{2}\dot{\phi}^2 + a^2 V(\phi)
    + \tilde{\rho}_{\dm}A^4(\phi) a^2 \\ \nonumber
    + &\rho_m a^2 + \rho_\Lambda a^2\ ,
\end{align}
where $\mathcal{H} = \dot{a}/a$, 
dots represent derivatives with respect to conformal time, and $\tilde{\rho}_{\dm}$ is the dark matter density in the Jordan frame. 
The equation of motion of the scalar is also modified in the Einstein frame and gains an additional source term dependent on the dark matter density  
\begin{equation}
    \ddot{\phi} + 2\mathcal{H}\dot{\phi} 
    = - a^2 V_{,\phi}
    - a^2 A_{,\phi}A^3(\phi) \tilde{\rho}_{\dm} \,.
\end{equation}
The homogeneous continuity equation for DM becomes 
\begin{align}
    \dot{\tilde{\rho}}_\dm 
    = -3 \left(\mathcal{H}+\frac{A_{,\phi}\dot{\phi}}{A}  \right)\tilde{\rho}_\dm \,,
\end{align}
that can be directly integrated giving
\begin{align}
    \tilde{\rho}_\dm 
    = \tilde{\rho}_\dm^{0} a^{-3}\left(\frac{A_0}{ A} \right)^3 \,,
\end{align}
where a subscript or superscript $0$ represents the value of a quantity today, at $a=1$. 
We hence define an auxiliary quantity, $\rho_{\dm}$, as 
\begin{align}
    \rho_\dm \equiv \tilde{\rho}_\dm A^3 \,,
    \label{eq:redefine_rho_dm}
\end{align}
which now dilutes like standard CDM as $\rho_{\dm} \propto a^{-3}$. 
This redefinition allows us to compute the effective potential $V_{\rm eff}(\phi)$ of the scalar in the Einstein frame as 
\begin{equation}
    V_{\rm eff}(\phi)
    = V(\phi)
    + A(\phi) \rho_{\dm} \,.
\end{equation}

The Hubble constraint equation evaluated today needs to be satisfied yielding 
\begin{align}
    3 M_P^2 \mathcal{H}^2_0 = \rho_m^0 +A(\phi_0) \rho_\dm^0 +\frac{1}{2}\dot{\phi}^2_0 +V_0 +\rho_\Lambda^0 \,,
\end{align}
which can be rewritten as 
\begin{align}
    1 = \Omega_m^0 +A(\phi_0) \Omega_{\dm,0} +\Omega_\phi^0 +\Omega_\Lambda^0 \,.
\end{align}
Note that the physical meaning of $\Omega_{\dm,0}$ is not the usual one and for all purposes, it is just an auxiliary variable. 
The relative DM gravitational density is given by $\Omega_{\dm,0} A(\phi_0)$, while the rest frame density is given by $\Omega_{\dm,0} A^{-3}(\phi_0)$.
Hence, at any point in time, the contribution of DM to the total energy budget of the Universe is $\rho_{\dm}A(\phi)$, which can be interpreted as a modulation of the mass of the DM particle. 

\section{Perturbation dynamics }
\label{sec:Perturbations}
We also account for the modified perturbation evolution of CEDE.
In this section, we follow the conventions of~\cite{Ma:1995ey}.
For convenience, we set up our perturbation calculations in the Einstein frame. 
For DM, we write the Einstein frame equations in terms of Jordan frame quantities, since the Jordan frame is the DM fluid rest frame where the physical interpretation of DM density perturbations is easier.
We then do a coordinate transformation on the relevant quantities of DM when determining their contributions to the total stress-energy tensor. 

The evolution of the density perturbation of DM depends on an additional source term due to the coupling to the scalar, while the DM velocity perturbation is now non-zero in the synchronous gauge and sourced by the scalar coupling: 
\begin{align}
\dot{\tilde{\delta}}_\dm 
    =& -\tilde{\theta}_\dm -\frac{1}{2}\dot{h} \nonumber\\
    &-3\left( \frac{A_{,\phi}}{A}\dot{\delta\phi} 
    + \dot{\phi} \delta\phi 
    \left( \frac{A_{,\phi\phi}}{A} - \frac{A_{,\phi}^2}{A^2} \right) \right) 
    \label{Eq:dm_continuity_pertb}
    \\
\dot{\tilde{\theta}}_\dm =& -\left( \mathcal{H} +{A_{,\phi}\over A}\dot{\phi}\right) \tilde{\theta}_\dm +{A_{,\phi}\over A} k^2 \delta\phi\,.
\end{align}
Note that these are the Einstein frame equations for the evolution of the rest frame (Jordan frame) DM density and velocity.

Likewise, with respect to the uncoupled scalar field model, the scalar field Klein-Gordon equation obtains additional source terms dependent on DM density:
\begin{align}
    \ddot{\delta\phi} 
    +2\mathcal{H}\dot{\delta\phi} &
    +k^2\delta\phi 
    +a^2V_{,\phi\phi}\delta\phi 
    +\dot{\phi}\frac{\dot{h}}{2} \nonumber\\
    =& -a^2 \left(A_{,\phi\phi} 
    +\frac{3A_{,\phi}^2}{A} \right) \rho_\dm \delta\phi \nonumber\\
    &- a^2 A_{,\phi} \rho_\dm \tilde{\delta}_\dm \,.
    \label{Eq:KG_pertb} 
\end{align}

Finally, we correct the contribution of DM density and velocity perturbations to the stress-energy tensor, accounting for the coordinate transformation between Einstein and Jordan frames:
\begin{widetext}
\begin{align}
    k^2\eta -\frac{1}{2}\mathcal{H} \dot{h} 
    &= - \frac{1}{2M_P^2} a^2 \left(\rho_{\rm m} \delta_{\rm m}  +4A_{,\phi} \rho_\dm \delta\phi +A\rho_\dm \tilde{\delta}_\dm +V_{,\phi}\delta\phi +\frac{\dot{\phi}}{a^2}\dot{\delta\phi} \right) \,,\nonumber\\
    k^2\dot{\eta} 
    &= \frac{1}{2M_P^2} a^2 \left[ (\rho_{\rm m}+P_{\rm m})\theta_{\rm m} +A\rho_\dm \tilde{\theta}_\dm +\frac{k^2}{a^2}\dot{\phi}\delta\phi \right] \,,\nonumber\\
    \ddot{h} +2\mathcal{H} \dot{h} -2k^2\eta 
    &= -\frac{3}{M_P^2}a^2 \left( \delta P_{\rm m} +\frac{\dot{\phi}}{a^2}\dot{\delta\phi} -V_{,\phi}\delta{\phi}\right) \,,\nonumber\\
    \ddot{h} +6\ddot{\eta} +2\mathcal{H}(\dot{h}+6\dot{\eta}) -2k^2 \eta 
    &= -\frac{3}{M_P^2} a^2 (\rho_{\rm m} +P_{\rm m}) \sigma_{\rm m} \,.
    \label{eq:einstein}
\end{align} 
\end{widetext}

\section{Initial conditions} \label{Sec:InitialConditions}
Independent of the choice of potential, the coupled scalar field is initially dominated by its kinetic energy $\dot{\phi}^2/2a^2$ 
and $\phi$ is rolling down the time-varying potential $A(\phi)\rho_\dm$. 
This holds for several order of magnitude in $A(\phi_{\rm i})$, as $\rho_\dm$ is very large at early times, and most reasonable choices of $\phi_{\rm i} \sim \Mpl$ and therefore $V(\phi_{\rm i}) \ll \rho_\dm(a_{\rm i})$ are not too large. 

During this time, there is little change in $\phi$ and its potential energy. 
At initial times, assuming $\ddot{\phi}_{\rm i}\rightarrow 0$, we set
\begin{align}
    2\mathcal{H}\dot{\phi_{\rm i}} 
    & \simeq - a^2 V_{,\phi}
    - a^2 A_{,\phi} \rho_{\dm} \nonumber\\
    \Rightarrow \dot{\phi}_{\rm i} &\simeq -\frac{a^2}{2\mathcal{H}} \big( V_{, \phi} + A_{,\phi}\rho_{\dm} \big) \,,
    \label{eq:phi_prime_init_cond}
\end{align}
which recovers the uncoupled regime as $A \rightarrow 0$. 
As $\dot{\phi}_{\rm i}$ is set deep in radiation domination, if the right-hand side is dictated by the DM term, $\dot{\phi}_{\rm i}$ is roughly constant, independent of $a$, validating setting $\ddot{\phi}_{\rm i} = 0$. 
Then, $\rho_{\rm scf} \propto a^{-2}$ at early times, dominated by the kinetic energy of $\phi$. 

The initial field location $\phi_{\rm i}$, on the other hand, is not set by an attractor solution, but is an input parameter that controls the maximal fractional energy density $f_{\ede}$ in CEDE. 
There is also a degeneracy between $\phi_{\rm i}$ and $\Omega_{\dm,{\rm i}}$ wherein changing $\phi_{\rm i}$ simply rescales $\Omega_{\dm,{\rm i}}$. 

The initial conditions for DM perturbations are left unchanged relative to the uncoupled case. 
At large scales, above the DM geodesic horizon, the DM fluid comoves with the synchronous gauge and hence synchronous gauge initial conditions apply to this scenario.
We set 
\begin{align}
    \tilde{\delta}_{\dm} &= \frac{3}{4} \delta_g \nonumber\\
    \tilde{\theta}_{\dm} &= 0 \,,
\end{align}
where $\delta_g$ is the photon density perturbation. 
The initial conditions for the scalar field perturbations are set as $\delta\phi \,, \dot{\delta\phi} = 0$, as they quickly evolve toward their attractor solutions. 

\section{Models of CEDE} \label{Sec:CEDEmodels}

With this general CEDE setup, we make the following choices of scalar field potential and form of the coupling 
\begin{align}
    A(\phi) &= e^{\beta \phi/\Mpl} \,\, {\rm and } \nonumber\\
    V(\phi) &= \lambda \phi^4 \,,
\end{align}
where $\lambda$ and $\beta$ are dimensionless constants, to illustrate CEDE dynamics. 
In presenting this toy model, we redefine the potential as 
\begin{align}
    V(\phi) &= \lambda_{\rm scf} V_0 \phi^4 \,,
\end{align}
where
\begin{align}
    V_0 & \equiv 2\times 10^{10} M_{\rm Pl}^{-2} H_0^2 \,,
\end{align}
complying with the CLASS unit convention. 
With this definition, $\lambda_{\rm scf} = 1$ and $\phi = \Mpl$ lead to a scalar field that dilutes close to matter-radiation equality. 

Of course, since the model contains no symmetry forbidding a mass term, the rules of effective field theory dictate that we should, 
in principle, include one, since if we do not then such a term will be generated by radiative corrections. 
For $f_\ede \sim O(10\%)$ and $z_c \sim z_{\rm eq}$, the scalar field must be ultra-light, with~$m_{\phi}(\phi_{\rm i}) \sim \sqrt{\lambda} \phi_{\rm i} \sim 10^{-28}~{\rm eV}$. 
The scalar mass will receive radiative corrections, for example from loops of dark matter, of order 
\begin{equation}
\delta m_{\phi} \sim \beta \frac{\Lambda^2}{M_{\rm Pl}}\,.
\end{equation}
The cutoff~$\Lambda$ should at the very least be $\gg$ than the dark matter particle mass~$m_{\rm dm}$, hence radiative stability requires~$m_{\rm dm} \ll {\rm eV}$. This will be the case if dark matter is axion-like. (A similar back-of-the-envelope shows that the quartic coupling is also radiatively stable for sufficiently light dark matter.) Thus the inclusion of a tree-level mass term, such that $V(\phi)$ becomes
\begin{align}
    V(\phi) = \frac{1}{2}m^2 \phi^2 + \lambda \phi^4 \, ,
\end{align}
with $m \lesssim 10^{-28}~{\rm eV}$ has little impact on cosmology. We leave the analysis of the model including a mass term to future work.

An uncoupled scalar field with a $\phi^4$ potential is initially frozen due to Hubble friction. 
When it begins to roll, its energy density dilutes $\propto a^{-4}$ when time-averaged over oscillations. 
Such a dissipation satisfies the EDE requirement of vanishing at late times and has already been explored in \cite{Agrawal:2019lmo}. 

\begin{figure*}
    \centering
    \includegraphics[width=0.49\textwidth]{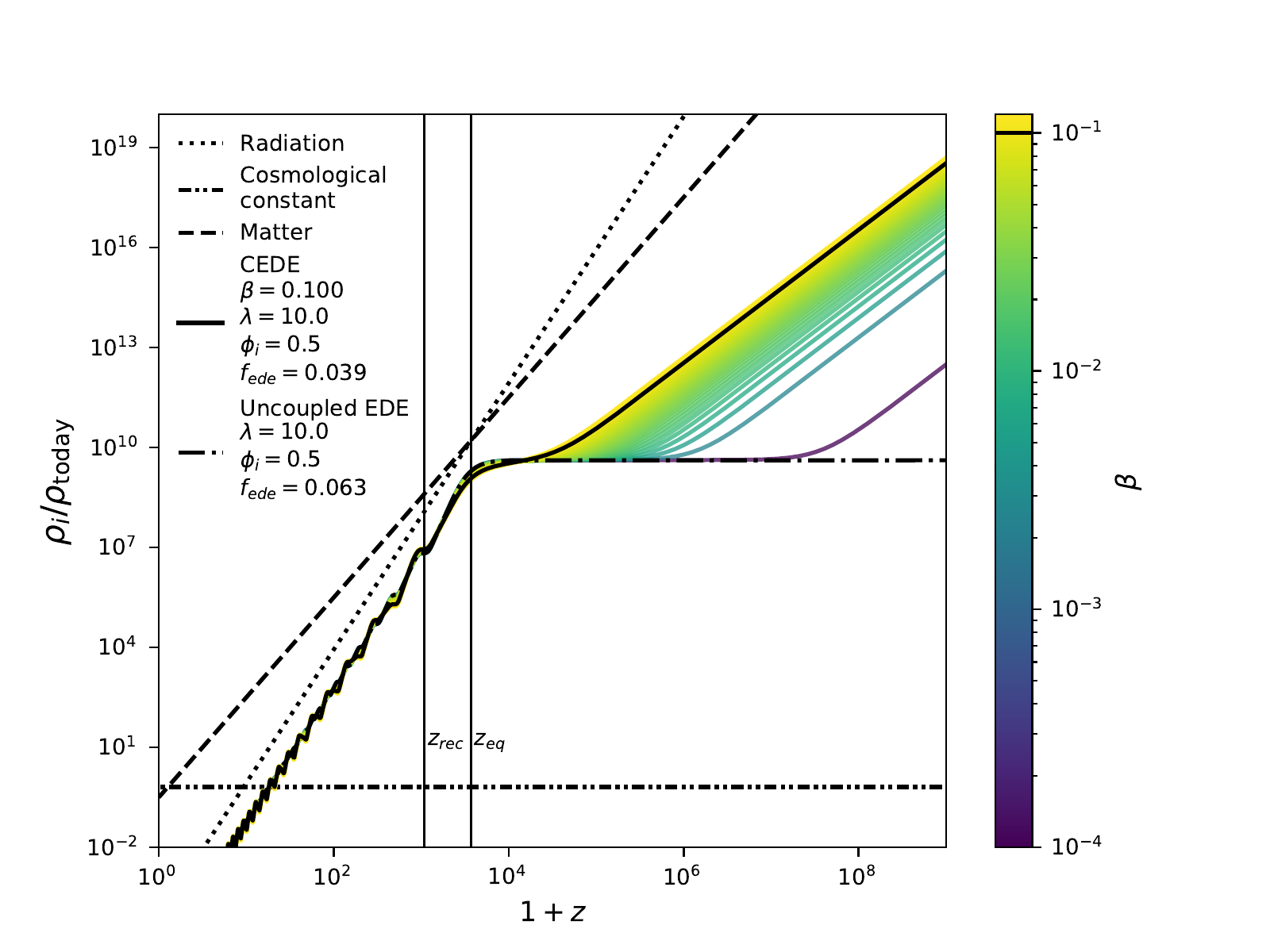}
    \includegraphics[width=0.50\textwidth]{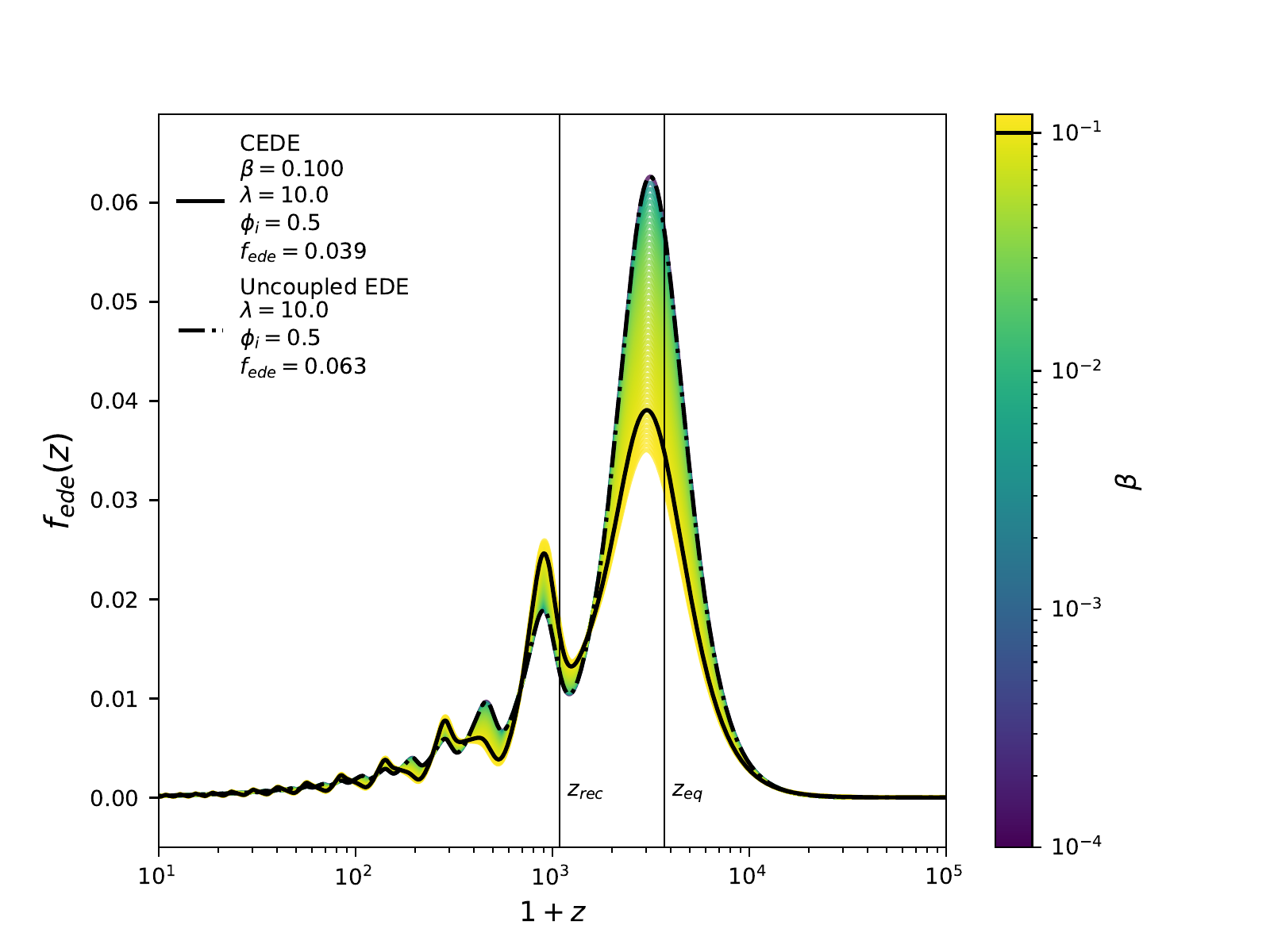}
    \includegraphics[width=0.49\textwidth]{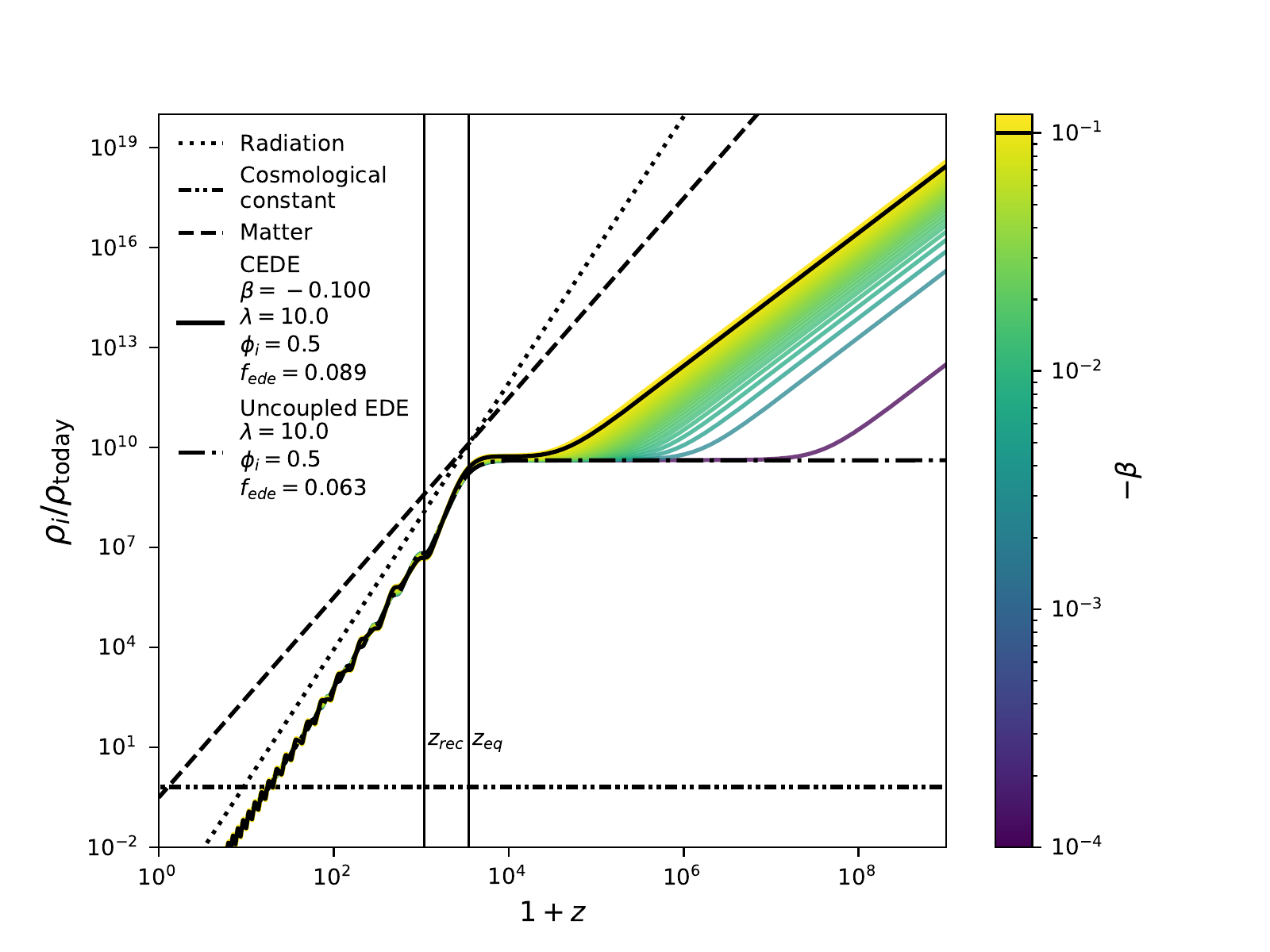}
    \includegraphics[width=0.50\textwidth]{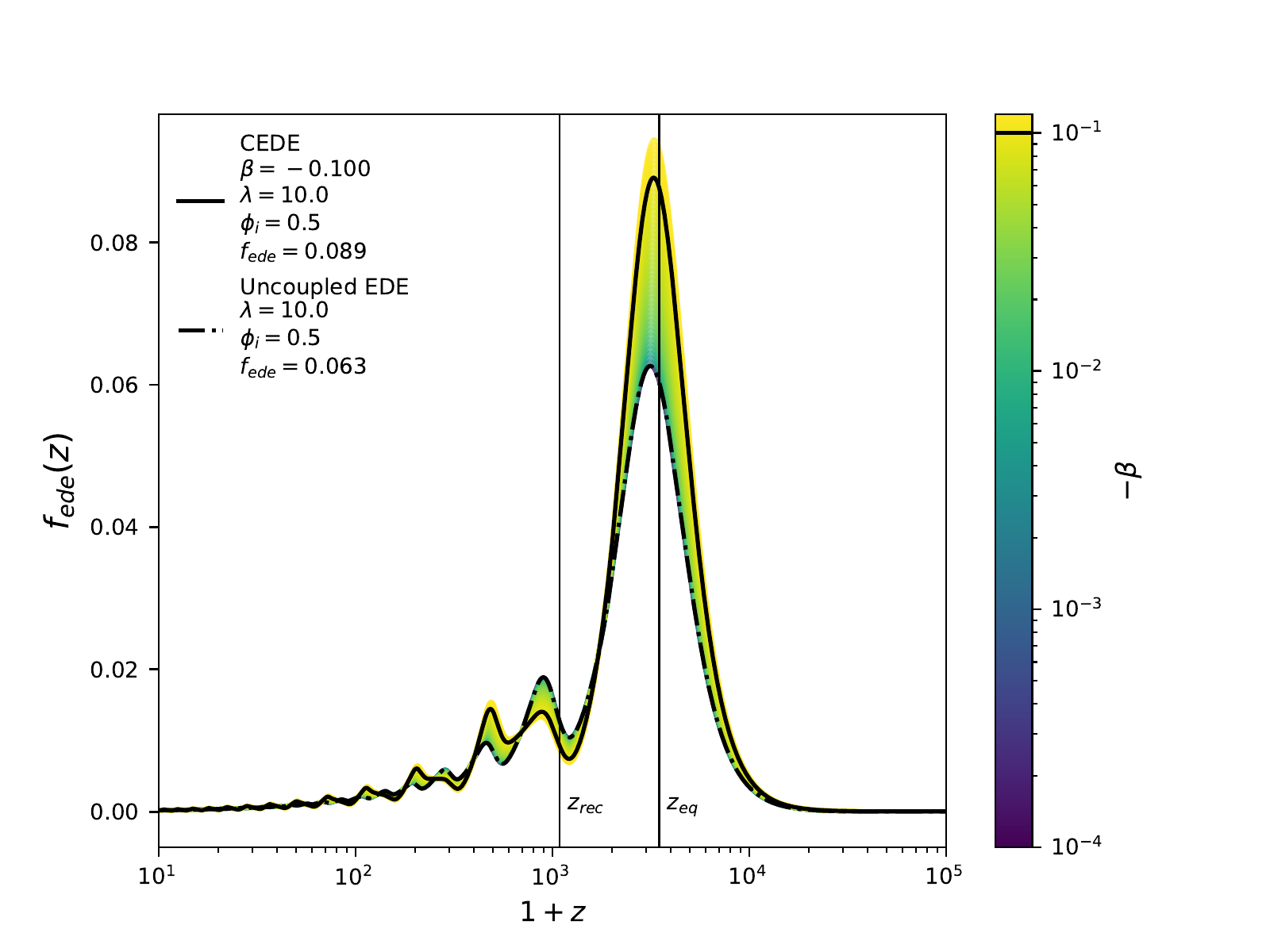}
    \caption{
    {\it Left:} Energy densities of various components of the Universe.
    We combine dark matter and baryons as the effect of CEDE on dark matter is unnoticeable at these scales. 
    The common parameters of uncoupled EDE (dot-dashed) and CEDE (solid) are $\phi_{\rm i} = 0.5 \Mpl$ and $\lambda = 10$. 
    In color, we show the variation of $\rho_{\rm scf}$ with $\beta$, for $\beta > 0$ in the top left and $\beta < 0$ in the bottom left panels. 
    The solid black curve sets $|\beta| = 0.1$. 
    {\it Right:} The fractional energy density $f_{\ede}$ in the EDE and CEDE scalar fields are shown for the same cosmologies.
    Depending on the sign of $\beta$, the DM interaction either increases its initial energy density and therefore $f_\ede$ (bottom right), or decreases $f_\ede$ as $\beta$ increases (top right). 
    \label{fig:beta_bg_behaviour}
    }
\end{figure*}

\begin{figure}
    \centering
    \includegraphics[width=0.49\textwidth]{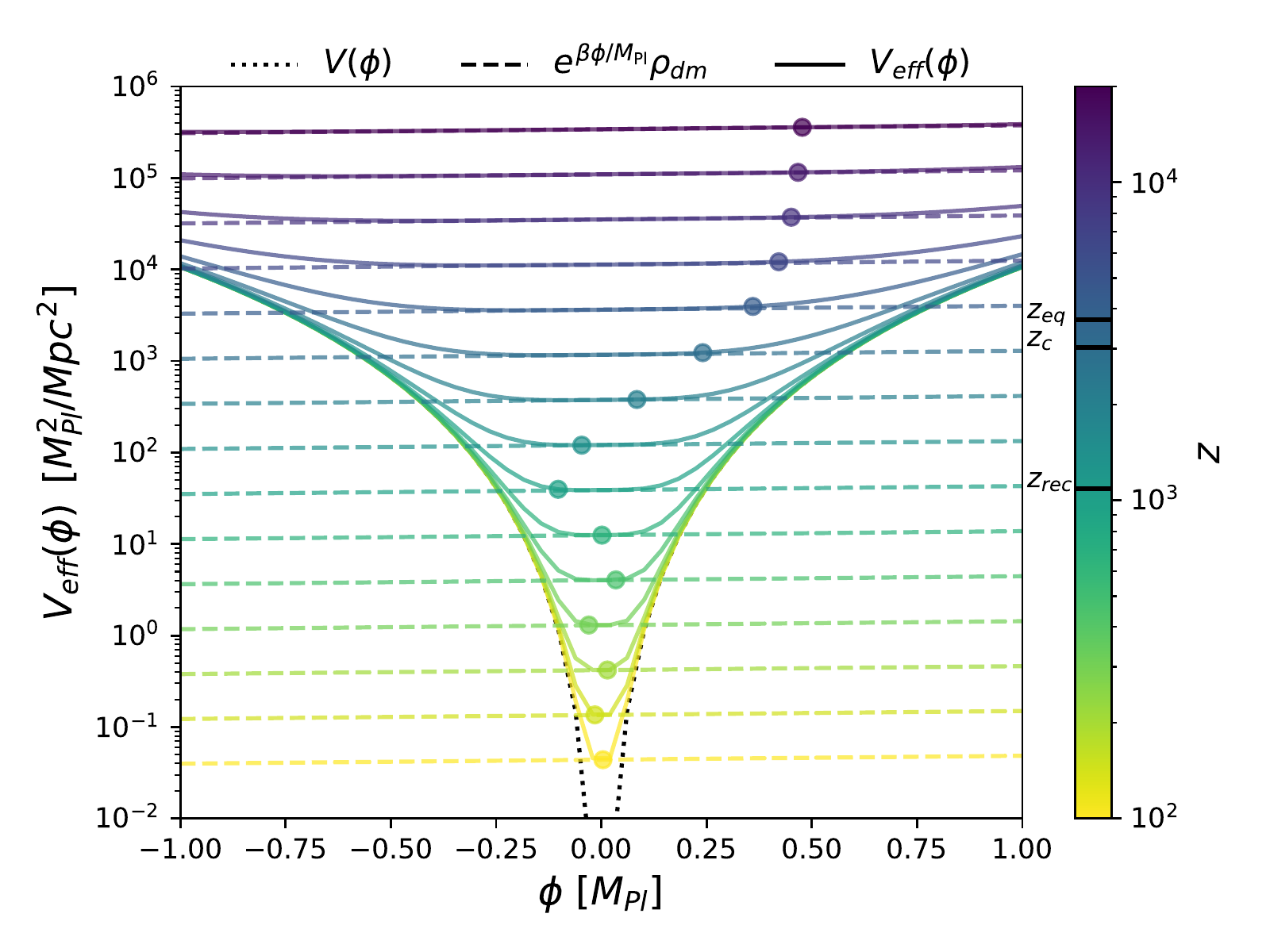}
    \caption{
    The effective potential $V_{\rm eff}(\phi)$ of the scalar (solid) shown in Eq.~\eqref{eq:eff_pot}, comprised of its potential $V(\phi)$ (dotted) and a source term $e^{\beta\phi/\Mpl}\rho_\dm$ (dashed) arising from its coupling to DM, for $\beta=0.1$. 
    Note that $e^{\beta\phi/\Mpl}\rho_\dm$ is not symmetric about $0$, but only resembles that in this zoomed-in plot. 
    The DM source term and the effective potential both depend on the DM density and are therefore time-dependent, and we show these quantities for several redshifts in color. 
    In dots, we show the position of $\phi$ for each of these redshifts, tracing its oscillations. 
    As $\rho_\dm$ decreases with redshift, $V_{\rm eff}(\phi) \rightarrow V(\phi)$. 
    Initially, $\phi_{\rm i} = 0.5 \Mpl$, shown by the topmost dot. 
    Then, when $V_{\rm eff}(\phi)$ becomes dominated by $V(\phi)$ in this region, between the third and fourth curves from the top, it begins to roll, with $z_c = 3026$. 
    \label{fig:V_eff}
    }
\end{figure}

At the background level, the introduction of the coupling to DM most noticeably modifies the early-time behavior of the scalar field, when the field is dominated by kinetic energy $\dot{\phi}^2/2a^2$, as described by Eq.~\eqref{eq:phi_prime_init_cond} with $\rho_{\rm scf}$ scaling as $a^{-2}$. 
This behavior can be seen at high redshifts in the left panels of Fig.~\ref{fig:beta_bg_behaviour}, produced using a modified version of the Boltzmann code CLASS \cite{Blas:2011rf}. 
Furthermore, as $\beta$ increases, the effective potential $V_{\rm eff}$ felt by the scalar differs from $V(\phi)$ to a greater extent, becoming increasingly asymmetric
\begin{equation}
    V_{\rm eff}(\phi) = \lambda \phi^4 
                        + e^{\beta \phi/\Mpl} \rho_{\dm} \,.
    \label{eq:eff_pot}
\end{equation}
It is this direct coupling to DM energy density that offers the possibility of EDE dynamics being triggered by DM becoming the dominant component of the Universe. 
In Fig.~\ref{fig:V_eff}, we show $V_{\rm eff}(\phi)$ as well as the field position as a function of redshift. 

The subsequent evolution of the field can be divided into two distinct scenarios - one in which $\beta \phi_{\rm i}/\Mpl < 0$ and another where $\beta \phi_{\rm i}/\Mpl > 0$.
Let us assume $\phi_{\rm i} / \Mpl >0 $. 
The more commonly explored chameleon dark energy case sets $\beta$ and $\phi_{\rm i}/\Mpl$ with opposite signs. 
In this scenario, the two terms contributing to $\dot{\phi}_{\rm i}$ in Eq.~\eqref{eq:phi_prime_init_cond} have opposite signs, slowing down $|\dot{\phi}_{\rm i}|$ relative to the $\beta >0$ case. 
Note that $|\dot{\phi}_{\rm i}|$ is still larger than in the uncoupled case, as the DM term dominates the contribution to $\dot{\phi}_{\rm i}$.  
Moreover, $\dot{\phi}_{\rm i} > 0$ and $\phi$ is initially being kicked up its native potential to larger values, moving towards the minimum of its effective potential in Eq.~\eqref{eq:eff_pot}, higher than in the uncoupled case. 
Then, as $|\beta|$ increases, the field is at a higher point in its potential at early times, with greater energy density than in uncoupled EDE or $\beta > 0$ CEDE.
The bottom right panel of Fig.~\ref{fig:beta_bg_behaviour} shows this most clearly.

On the other hand, for $\beta \phi_{\rm i}/\Mpl >0$, $|\dot{\phi}_{\rm i}|$ is larger than in the $\beta < 0$ CEDE case, and is negative. 
Hence, $\phi$ is lower in its potential relative to the uncoupled or $\beta < 0$ CEDE cases at early times. 
Accordingly, as $\beta$ increases, CEDE has smaller $f_\ede$ in this case than in uncoupled EDE, most evident in the top right panel of Fig.~\ref{fig:beta_bg_behaviour}. 

As DM dilutes away, the native potential $V(\phi)$ of the scalar begins to dictate its dynamics. 
Both $\beta >0$ and $\beta<0$ scenarios may then become Hubble frozen for some decades in redshift.
The value of $\beta$ controls the duration of this Hubble-frozen period, with higher $\beta$ leading to a smaller or no frozen window.
For smaller values of $\beta$, it is $\lambda_{\rm scf}$ that controls the redshift $z_c$ at which $f_\ede$ peaks, similar to uncoupled EDE. 

The scalar then begins to roll and oscillate in $V_{\rm eff}$ about a new, time-varying minimum shifted from 0, defined by the solution to 
\begin{equation}
    4\lambda \phi_{\rm min}^3 = - \frac{\beta}{\Mpl}e^{\beta \phi_{\rm min}/\Mpl} \rho_{\dm} \,.
\end{equation}
Although the time-averaged density of the field falls as $a^{-4}$ during this period, $\phi$ undergoes asymmetric oscillations about this new minimum. 
This shifts the odd (even) peaks in $f_\ede$ to lower (higher) energy density than in the symmetric-potential uncoupled EDE case for $\beta > 0$ and vice versa for $\beta < 0$. 
The right panels of Fig.~\ref{fig:beta_bg_behaviour} show these shifted oscillations. 

\section{Impact of non-zero $\beta$} \label{Sec:BetaResults}

\begin{figure*}
    \centering
    \includegraphics[width=0.49\textwidth]{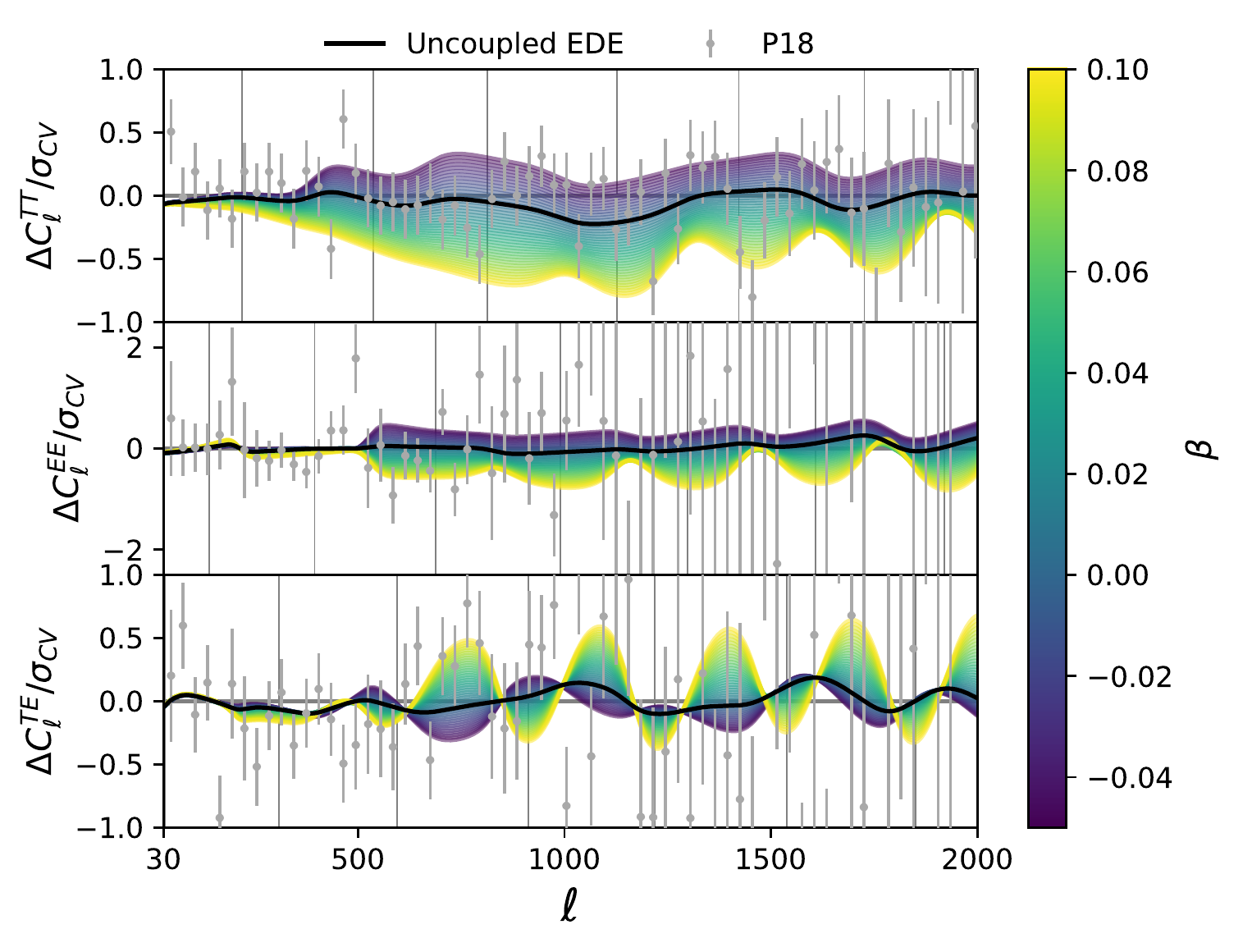}
    \includegraphics[width=0.49\textwidth]{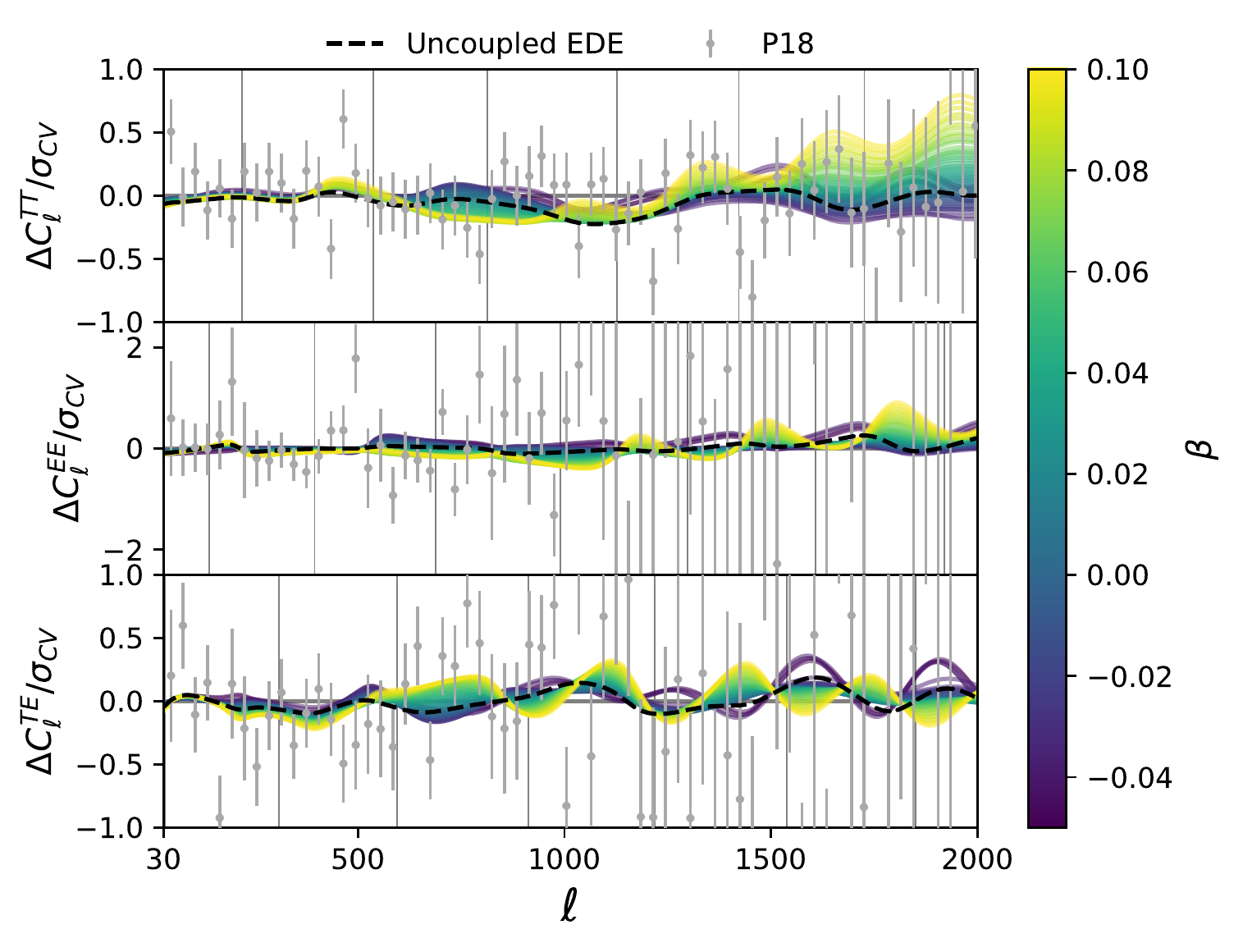}
    \caption{
    The impact of varying $\beta$ on CMB residuals. 
    The dashed black curve shows the maximum likelihood (ML) uncoupled EDE relative to a \LCDM\ fit to data. 
    The gray points are Planck 2018 data points and the vertical lines mark the location of the peaks in each spectrum. 
    In the colored curves, we begin from the ML uncoupled EDE point and linearly tune $\beta \in [-0.05, 0.1]$. 
    {\it Left:} All other cosmological parameters are held fixed at their uncoupled EDE ML values. 
    {\it Right:} We optimize over the other cosmological parameters as we tune $\beta$. 
    Hence, for any given $\beta$, the right plot maximizes the likelihood. 
    }
    \label{fig:tune_beta_cls}
\end{figure*}

In Fig.~\ref{fig:tune_beta_cls}, we present the impact of tuning $\beta$ around the maximum likelihood point of uncoupled EDE on CMB residuals, either keeping all other cosmological parameters fixed, in the left panel, or optimizing them to maximize the likelihood at each value of $\beta$, in the right panel.
Here, uncoupled EDE and a reference \LCDM\ model are fit to Planck 2018 CMB temperature, polarization and lensing spectra \cite{Aghanim:2018eyx, Aghanim:2019ame,Aghanim:2018oex}, baryon acoustic oscillation (BAO) data \cite{Alam:2016hwk, Ross:2014qpa, Beutler:2012px}, Pantheon supernova data \cite{Scolnic:2017caz} and the local $H_0$ measurement \cite{Riess:2019cxk} with a Markov chain Monte Carlo (MCMC) algorithm \cite{Lewis:2013hha,Lewis:2002ah, Neal:2005,gelman1992} using the sampler Cobaya \cite{Torrado:2020dgo} and analysis code GetDist \cite{Lewis:2019xzd}. 
The ML points are found using the minimizer algorithm BOBYQA \cite{BOBYQA,2018arXiv180400154C, 2018arXiv181211343C}. 

As $\beta \rightarrow 0$ while other cosmological parameters are optimized, the CEDE residuals begin to resemble the ML uncoupled EDE residuals. 
By comparing the two figures, we can gauge what effects, due to non-vanishing $\beta$, can be reabsorbed by a change in other parameters.
In contrast, features that do not change between the panels are the ones that capture the improvement that $\beta$ provides to the fit. 
The two panels show marked differences in all three spectra, but most so in TT. 
The changes in the EE spectra are nearly entirely reabsorbed by changes in other cosmological parameters. 
Most notably, the figures show how modulating $\beta$ can improve the fit to CMB TT and TE spectra in the $500 < \ell < 1500$ range, following the oscillatory behavior of \LCDM\ CMB residuals. 

To fully and robustly understand the impact of data on CEDE models, we will perform Markov chain Monte Carlo (MCMC) searches in parameter space in future work. 
Here, we simply highlight the interesting features of such models and the expectation of improvement over an uncoupled EDE model. 

\section{Conclusions} \label{Sec:Conclusions}

The concordance \LCDM\ model fits numerous cosmological observations exceedingly well. 
Its most notable achievement with a single dataset is the fit to the complicated cosmic microwave background power spectra with just 6 independent parameters \cite{Aghanim:2018eyx}. 
Equally impressive is the fact that the same model fits a diverse set of measurements in the late-time universe, including probes of the expansion history and the growth of structure \cite{Alam:2016hwk,Ross:2014qpa,Beutler:2012px,Scolnic:2017caz,Abbott:2021bzy,Muir:2020puy}. 
As measurement precision has improved, while most cosmological data remain individually consistent with \LCDM, tensions have emerged when comparing constraints on common parameters from the CMB versus late time observations. 
Of these, the Hubble tension has received the greatest attention from observers and theorists alike and offers exciting hints of physics beyond the phenomenological \LCDM\ model \cite{Verde:2019ivm,Freedman:2017yms,DiValentino:2020zio}.
Add to this the developing large-scale structure tension and physicists are presented with cracks in \LCDM\ through which to explore the dark sector \cite{DiValentino:2021izs,DiValentino:2020vvd}. 

Early dark energy models aim to do exactly this -- they postulate new physics in the dark sector  to resolve the Hubble tension without compromising the fit to CMB, BAO or supernovae data which are fit well by \LCDM. 
The common features in these models are (i) a pre-recombination energy injection close to matter-radiation equality, (ii) rapid dilution of energy density thereafter such that the impact of EDE is localized in redshift and (iii) preference for a higher dark matter density $\omega_\dm$  when fit to the CMB, which generally worsens the LSS tension \cite{Vagnozzi:2021gjh,Jedamzik:2020zmd,Ivanov:2020ril,Hill:2020osr}. 
These features raise several questions, including how such an EDE might theoretically arise, a new `why then' problem akin to the `why now' problem of late-time dark energy, and concerns over the exacerbation of the LSS tension. 
While several theoretical models for EDE have been proposed \cite{Poulin:2018dzj,Smith:2019ihp,Berghaus:2019cls,Niedermann:2019olb,Agrawal:2019lmo,Sakstein:2019fmf,Alexander:2019rsc,Freese:2021rjq,Ye:2020btb,Braglia:2020bym,Gogoi:2020qif,Garcia:2020sjl}, few of these provide answers to the `why then' problem and none have simultaneously resolved the Hubble and LSS tensions. 

In this paper, we suggest addressing both the latter questions with ``chameleon early dark energy'' (CEDE), inspired by chameleon models of late-time dark energy \cite{Brax:2004qh}. 
We introduce a scalar field 
that conformally couples to dark matter. 
This set up provides grounds for tying the dynamics of EDE to the onset of matter domination at $z_{\rm eq}$. 
Moreover, interactions between a scalar field and dark matter may also have implications for the clustering of matter and hence the LSS tension.
Here, we have specifically explored the scenario wherein $V(\phi) = \lambda \phi^4$ and the coupling takes the form $e^{\beta \phi/\Mpl}$, where $\lambda$ and $\beta$ are dimensionless, 
but note that CEDE models with different potentials and couplings will have variable impact on data. 
An uncoupled EDE model with a $\phi^4$ potential has been studied in the literature \cite{Agrawal:2019lmo}, but is not very successful at resolving the Hubble tension while simultaneously providing as good a fit to data as \LCDM. 
As we have mentioned, more complicated EDE potentials, such as $V(\phi)\sim (1 - \cos (\phi) \,)^n$, can provide a better fit to the data, but at the expense of theoretical challenges. 
The introduction of CEDE allows us to ameliorate the issue of theoretical fine tuning, while providing another parameter to tune to better fit data.
In particular, in exploring a $\phi^4$ potential for CDE, this is achieved with the addition of a single new adjustable parameter, $\beta$. 

At the background level, this chameleon coupling alters the initial field velocity $\dot{\phi}_{\rm i}$ and makes the effective potential asymmetric. 
The field dilutes as $a^{-2}$ initially, then becomes Hubble frozen before thawing and oscillating in its asymmetric potential, diluting as $a^{-4}$ when averaged over oscillations. 
In the preferred regions of parameter space, the impact on dark matter at the background level is minimal, effectively captured by a modulation of the DM particle mass. 
As dark matter behaves like CDM at late times, we can justify using non-linear codes written for \LCDM\ when studying CEDE. 
At the perturbation level, besides effects common with uncoupled EDE, the leading order impact to the gravitational potentials $\Phi+\Psi$ comes from the perturbation of the temporal modulation of the DM gravitational mass, i.e. the terms dependent on $\tilde{\delta}_\dm$ and $\tilde{\theta}_\dm$ in Eq.~\eqref{eq:einstein}. 

The EDE model that is currently most successful at resolving the Hubble tension has a finely-tuned and difficult-to-motivate potential. 
On the data end, recent high-$\ell$ CMB polarization results from ACT have demonstrated a preference for non-zero EDE at over $3\sigma$ \cite{Poulin:2021iyx, Hill:2021yec}. 
As observers have not yet reached a consensus on the Hubble tension \cite{Freedman:2019jwv, Riess:2020fzl} and in light of new data that prefers EDE, it is vital to build models resembling EDE that are well-motivated by theory and can be tested with upcoming CMB polarization data from ACT and SPT \cite{Lin:2020jcb,Chudaykin:2020igl,Chudaykin:2020acu,Smith:2019ihp}. 

Lastly, we note that the recent DES Y3 results indicate a lower tension with Planck \cite{Abbott:2017wau,Krause:2017ekm,Abbott:2021bzy}, which lowers the cumulative LSS tension level. 
This lower $S_8$ tension level for lensing suggests the jury is still out on the need for new physics to explain the growth of structure. 
Spectroscopic surveys of galaxies provide another probe of structure via galaxy motions -- redshift space distortions in the measured galaxy power spectrum. 
Again, the tension with Planck is at a similar level, as discussed e.g. by \citet{Nunes:2021cin}, it is not at high statistical significance. 
Joint analyses of weak lensing, redshift space distortions and other measures of large scale structure may offer new guidance, and  upcoming galaxy surveys with the Dark Energy Spectroscopic Instrument, Euclid and the Rubin Observatory  \citep{DESI_Part_1, Abell:2009aa,Laureijs:2011gra} will almost certainly produce tighter constraints on $\sigma_8$ and other late-universe parameters. 
As these tensions continue to develop, theoretical solutions that use these anomalies to probe physics beyond the \LCDM\ model are vital to our understanding of the Universe. 

\acknowledgements{
We are grateful to Wayne Hu, Meng-Xiang Lin, Vivian Poulin, Marc Kamionkowski and Jeremy Sakstein for helpful discussions. 
We are particularly thankful for Tristan Smith's inputs on a previous version. 
This work is supported in part by NASA ATP grant 80NSSC18K0694. TK and MR are also supported in part by funds provided by the Center for Particle Cosmology. MT and JK are also supported in part by US Department of Energy (HEP) Award DE-SC0013528. 
}


\bibliographystyle{apsrev4-1}
\bibliography{paper}
\end{document}